\documentclass[preprint,showkeys,showpacs]{revtex4}

\usepackage{graphicx}

\begin{document}

\title{Stability of thermal structures
       with an internal heating source}

\author{N\'estor S\'anchez}
\email{nestor@iaa.es}
\affiliation{Instituto de Astrof\'{\i}sica de Andaluc\'{\i}a,
             CSIC, Granada, Spain.}
\author{Eugenio L\'opez}
\affiliation{Departamento de F\'{\i}sica y Matem\'aticas,
             Universidad Nacional Experimental Francisco
             de Miranda, Coro, Venezuela.}

\begin{abstract}
{\bf Abstract:}
We study the thermal equilibrium and stability of isobaric,
spherical structures having a radiation source located at its
center. The thermal conduction coefficient, external heating and
cooling rates are represented as power laws of the temperature.
The internal heating decreases with distance from the source
$r$ approximately as $\sim \exp(-\tau)/r^2$, being $\tau$ the
optical depth. We find that the influence of the radiation
source is important only in the central region, but its effect
is enough to make the system thermally unstable above a certain
threshold central temperature. This threshold temperature
decreases as the internal heating efficiency increases, but,
otherwise, it does not depend on the structure size. Our
results suggest that a solar-like star migrating into a
diffuse interstellar region may destabilize the surrounding
medium.\\
\begin{center}
{\bf Estabilidad de estructuras t\'ermicas
con una fuente interna de calentamiento.}
\end{center}
{\bf Resumen:}
En este trabajo estudiamos el equilibrio t\'ermico y la estabilidad
de estructuras isob\'aricas y esf\'ericas que tienen una fuente
de radiaci\'on en su centro. La conducci\'on t\'ermica y las tasas
de calentamiento externo y enfriamiento son representadas como
leyes de potencia de la temperatura. El calentamiento interno
disminuye con la distancia a la fuente $r$ aproximadamente como
$\sim \exp(-\tau)/r^2$, siendo $\tau$ la profundidad \'optica. Los
resultados indican que la influencia de la fuente de radiaci\'on
es importante solamente en la regi\'on central, pero su efecto
es suficiente para que el sistema se vuelva t\'ermicamente
inestable por encima de una cierta temperatura umbral en
el centro. Esta temperatura umbral disminuye a medida que
la eficiencia del calentamiento interno aumenta, pero por
otro lado la misma no depende del tama\~no de la estructura.
Nuestros resultados sugieren que una estrella de tipo solar
que migre a trav\'es del medio interestelar difuso puede
llegar a desestabilizarlo.
\end{abstract}


\maketitle

\section{Introduction}

Most of the structures observed in Astrophysics can be
fully explained only if one takes into account several
physical processes simultaneously, such as dynamics,
self-gravitation, magnetic fields, radiation transfer,
thermal conduction, complex chemical networks, etc.
Since the seminal paper by Field \cite{Fie65}, a lot of
work has been devoted to the study of the thermal and
mechanical equilibrium of structures under the action
of different physical mechanisms \cite{Mee96}.
One of the motivations behind this kind of studies has
been to explain the formation of structures via thermal
instabilities. Nowadays, it is clearly established that
thermal instabilities play a major role in several
astrophysical contexts, such as solar chromosphere and
corona, interstellar medium, planetary nebulae, and
intergalactic medium \cite{Spi78,Fer01,Mee96,Dib04,Sti06}.
A full modeling of these systems may be a relatively difficult
task requiring large computational resources (e.g. see Refs.
\cite{Sly05,Pio05,Vaz06}). On the contrary, the development
of simple models including few physical processes allows
to clarify the specific role of each process, setting
up the framework for more realistic models.

When only thermal conduction and a net heat-loss function are
considered, a variety of (isobaric) thermal structures are
obtained and stability criteria can be derived analytically
\cite{Iba92}. These criteria can be extended analytically up
to the second order \cite{Iba93}. In spite of the simplicity
of this approach (many physical processes are not considered)
these results can be applied successfully to the atomic phase
of the ISM \cite{Par93}, molecular clouds \cite{Iba94}, and
the solar corona \cite{Men97}. Usually, it is assumed that
the heating mechanisms are ``local" functions, in the sense
that they depend only on the local thermodynamical state
(temperature, density). However, two factors may complicate
this picture. On the one hand, the heating mechanism could be an
explicit function of the position in the structure. This situation
applies, for example, if a heating source is located close to the
structure, such as an interstellar region with a nearby star (or
star cluster), or the solar corona heated by the chromosphere. For
coronal loops, it has been shown that the existence or not of
thermal equilibrium solutions and their stability is subordinated
to the spatial dependence of the heating \cite{Men97}. On the
other hand, if the heating mechanism is associated with radiative
processes, the opacity adds a non-local character to the heating
function making it difficult to perform an analytical study.
In Ref. \cite{Par96} the stability of structures heated by
radiative processes was analyzed. Generally, the results
showed that the effect of increasing the opacity is to
increase the stability of the thermal structures.
However, in order to achieve certain analytical
results that work was limited to the unrealistic
case of a slab-like configuration irradiated
simultaneously on both sides.

In the present paper, our attention is focused on investigating
the thermal equilibrium and the stability of spherical structures
heated by a radiation source located at its center. Besides the
pure academic interest of generalizing the previous results
\cite{Iba92,Iba93,Par93,Iba94,Par96}
by including an internal heat source, this kind of study
could be very useful because this is a common configuration, in
particular in Astrophysics. We try to keep the approach as general
as possible and, therefore, our results can in principle be
directly applied to any situation in which an isobaric, spherical
structure has an internal radiation source. However, we use values
for constants and parameters that are appropriate for typical
interstellar conditions because in the future we plan to
generalize this work in order to be applicable to interstellar
regions in the Galaxy. It is known, for example, that the diffuse
ISM is linearly thermally stable but nonlinearly
unstable (e.g. \cite{Hen99}).
Moreover, it is expected that stars
systematically pass through the ISM. Thus,
it is relevant to study whether these stars can sufficiently
perturb an interstellar region as to induce thermal instability.
Obviously, many more physical mechanisms have to be considered
to describe this situation adequately but, as mentioned before,
in a first approximation this study can help to understand the
role played by this type of realistic heating.
In Sec.~\ref{sec_ecuaciones}
we present the basic equations, and in Sec.~\ref{sec_criterio}
we briefly explain the stability criterion we are using. The
effect of the internal source on the stability is discussed in
Sec.~\ref{sec_resultados} and, finally, the main conclusions are
summarized in Sec.~\ref{sec_conclusiones}.

\section{\label{sec_ecuaciones} Basic equations}

The equation of energy conservation for a static, thermally
conducting fluid, with pressure $p$ constant in time $t$ can
be written in the form \cite{Lan87}:
\begin{equation}
\label{energia1}
\rho c_p \frac{\partial T}{\partial t} =
\nabla \cdot \left[ \kappa (\rho,T) \nabla T \right] +
\Gamma (\rho,T) + \Lambda (\rho,T)\ ,
\end{equation}
where $T$ is the temperature, $\rho$ the density, $c_p$ the
specific heat per unit mass, and $\kappa (\rho,T)$ is the thermal
conduction coefficient. The functions $\Gamma (\rho,T)$ and
$\Lambda (\rho,T)$ represent the heating and cooling rates per
unit volume, respectively. In many astrophysical situations the
quantities $\kappa$, $\Gamma$ and $\Lambda$
can be approximated as being proportional to power laws
of the form $\sim \rho^{\alpha} T^{\beta}$ \cite{Iba91}. Under
the assumption of isobaricity, we can eliminate the density as
an independent variable and write
\begin{equation}
\label{indices}
\kappa=\kappa_0 T^k\ ,\
\Gamma=\Gamma_0 T^m\ ,\
\Lambda=\Lambda_0 T^n\ ,
\end{equation}
where the constants $\kappa_0$, $\Gamma_0$, $\Lambda_0$ and the
indices $k$, $m$ and $n$ are given by the physical processes
involved. For instance, a typical interstellar region in the
range of temperature $10^2 \alt T \alt 10^4\ K$ with thermal
conduction by neutral particles, heating proportional to the
density, and cooled by collisions between particles, may be
characterized by $k=1/2$, $m=-1$ and $n=-3/2$ \cite{Par93,Iba91}.

In this work we consider an additional heating mechanism. Let
us assume a spherically symmetric region
with radius $R_b$ that has a radiation source located at
its center. The photons coming from this source can heat the
surrounding gas. In the ISM this heating is mainly because
the UV radiation field is able to photoeject electrons from
atoms and dust which subsequently thermalize through collisions
\cite{Spi78}. The energy absorbed per unit volume is \cite{Hol90}:
\begin{equation}
\Gamma_s(r) = \gamma \sigma n F(r) \ ,
\end{equation}
where $F(r)$ is the radiation flux coming from the central
source, $n$ is the number of atoms per $cm^3$, $\sigma$ is
the photon capture cross section, and the efficiency
parameter $\gamma$ is the fraction of incident energy
which is actually absorbed by the gas. The radiation flux
decreases with distance from the center ($r$) as $1/r^2$.
In order to avoid dealing with divergences at $r=0$ when
integrating Eq.~\ref{energia1}, we impose a spatial
smoothing scale $R_s$. Thus, the radiation flux can be
represented in the form \cite{Mih84}:
\begin{equation}
\label{flujo}
F(r) = \frac{F(0)\exp{(-\tau)}}{\left( 1+(r/R_s)^{2a}
\right)^{1/a}} \ ,
\end{equation}
where $\tau$ is the total optical depth from the central
source to $r$:
\begin{equation}
\label{tau}
\tau = \int_{0}^{r} \sigma n dr \ .
\end{equation}
The value of the arbitrary constant $a$ determines how quickly
the behavior changes from $\sim F(0)$ when $r \ll R_s$ to $\sim
1/r^2$ when $r \gg R_s$. We have kept $a$ unchanged ($a=10$),
but several tests showed that our results do not depend on $a$
as long as the condition $R_s \ll R_b$ remains fulfilled.
The flux at the center can be calculated by assuming emission
from a black body of effective temperature $T_0$:
\begin{equation}
F(0)=\pi \int_{UV} B_{\nu} (T_0) d\nu\ ,
\end{equation}
$B_{\nu}$ being the Planck's function. The integration is done
in the far ultraviolet ($900-1100\ \mathring{A}$) because
this is the range in which photons heat the ISM efficiently
\cite{Bak94,Wol95}.

With these considerations, Eq.~\ref{energia1} can be
rewritten as
\begin{eqnarray}
\label{energia2}
\nonumber
\rho c_p \frac{\partial T}{\partial t} & = &
\frac{1}{r^2} \frac{\partial}{\partial r}
\left( r^2 \kappa_0 T^k \frac{\partial T}{\partial r} \right) +
\Gamma_0 T^m - \Lambda_0 T^n \\
& & +
\frac{\gamma \sigma n F(0)\exp(-\tau)}{\left( 1+(r/R_s)^{2a}
\right)^{1/a}} \ .
\end{eqnarray}
This is the integro-differential equation whose stationary
solutions (and their stability) we want to study in this
work. It is convenient, however, to express it in terms of
some non-dimensional quantities. If $T_{eq}$ is the temperature
to which $\Gamma = \Lambda$, i.e.:
\begin{equation}
T_{eq} = (\Gamma_0 / \Lambda_0)^{1/(n-m)} \ ,
\end{equation}
then we can define a non-dimensional temperature as:
\begin{equation}
\label{tem1}
\theta = T/T_{eq}\ .
\end{equation}
Given the (external) heating and cooling mechanisms, $T_{eq}$ is
a constant representing the temperature at the thermal equilibrium
when there exists no internal heating ($\Gamma_s = 0$). We also
write the position normalized to the region size:
\begin{equation}
\label{tem2}
z = r/R_b\ ,
z_s = R_s/R_b\ .
\end{equation}
We have kept the smoothing length fixed at
the arbitrary value $z_s = 10^{-8}$, but,
as mentioned before, our results do not depend on the
exact $z_s$ value as long as $z_s \ll 1$ is satisfied.
The total optical depth from the center to the boundary
for the particular case $T=T_{eq}=constant$ is:
\begin{equation}
\label{tem3}
\tau_0 = \frac{\sigma (p/k_B) R_b}{T_{eq}} \ ,
\end{equation}
where the ideal gas equation of state $p/k_B = nT$
(being $k_B$ the Boltzmann's constant) has been used.
Therefore, including Eq.~\ref{tem1} to \ref{tem3} and
normalizing the flux at the center to the equilibrium flux,
i.e. $\tilde{F} = F(0)/F_{eq}$ where $F_{eq}$ is the black
body flux calculated using $T=T_{eq}$, we see that the
stationary solutions of Eq.~\ref{energia2} are the solutions of:
\begin{eqnarray}
\label{estacionaria}
\nonumber
\frac{1}{z^2} \frac{d}{dz} \left( z^2 \theta^k \frac{d\theta}{dz}
\right) + \lambda \left( \theta ^m - \theta ^n \right) + & \\
\frac{\eta \ \tilde{F}
\exp\left[-\tau_0 \int_0^z (1/\theta)dz\right]}{\theta
\left( 1+(z/z_s)^{2a} \right)^{1/a}} & = 0\ ,
\end{eqnarray}
where
\begin{equation}
\label{lambda}
\lambda = \frac{\Gamma_0 R_b^2}{\kappa_0 T_{eq} ^{k-m+1}} \ ,
\end{equation}
and
\begin{equation}
\label{eta}
\eta = \frac{\sigma (p/k_B) F_{eq} \gamma R_b^2}
{\kappa_0 T_{eq} ^{k+2}} \ .
\end{equation}
To solve Eq.~\ref{estacionaria} we need two boundary
conditions for the temperature. This paper is aimed to
quantify the effects of an internal heat source on
the stability of the steady structures taking as
reference the previous papers for which the stability
criteria are known. Then, for convenience, we assume
exactly the same boundary conditions used in Refs.
\cite{Iba92,Iba93,Par93,Iba94,Par96,Iba91}:
\begin{equation}
\label{frontera1}
\frac{d\theta}{dz} = 0\ \ {\rm at}\ \ z=0 \ ,
\end{equation}
\begin{equation}
\label{frontera2}
\theta = \theta_b\ \ {\rm at}\ \ z=1 \ .
\end{equation}
The first condition follows from the symmetry of the
configuration: the resulting temperature distribution
from the center $z=0$ to the $z=+1$ side should be
identical to the $z=-1$ side. This is the only condition
compatible with this property if we impose the additional
physically motivated requirements that both the temperature
and its derivative exist and be finite.
The second condition, i.e. fixed boundary temperature
$\theta_b$, is necessary in order to integrate the energy
equation. We write down explicitly this condition (fixed
temperature) at $z=1$ only to be consistent with the
previous papers, but actually a fixed value at any
other position would be equally valid as long as the
integration over $z$ can be performed. In fact, we are
assuming a fixed central temperature and the integration
of Eq.~\ref{estacionaria} is performed outwards from this
central position (see Sec.~\ref{sec_resultados}).
This is equivalent to converting the two point boundary
value problem (which requires considerably more numerical
effort) into an initial value problem. In practice there is
no difference between both approaches because the result
of interst for this work is the entire family of solutions
rather than a single solution and, therefore, we have to
span the whole range of possible central (and boundary)
temperatures.

\section{\label{sec_criterio} Stability criterion}

\begin{figure*}
\includegraphics{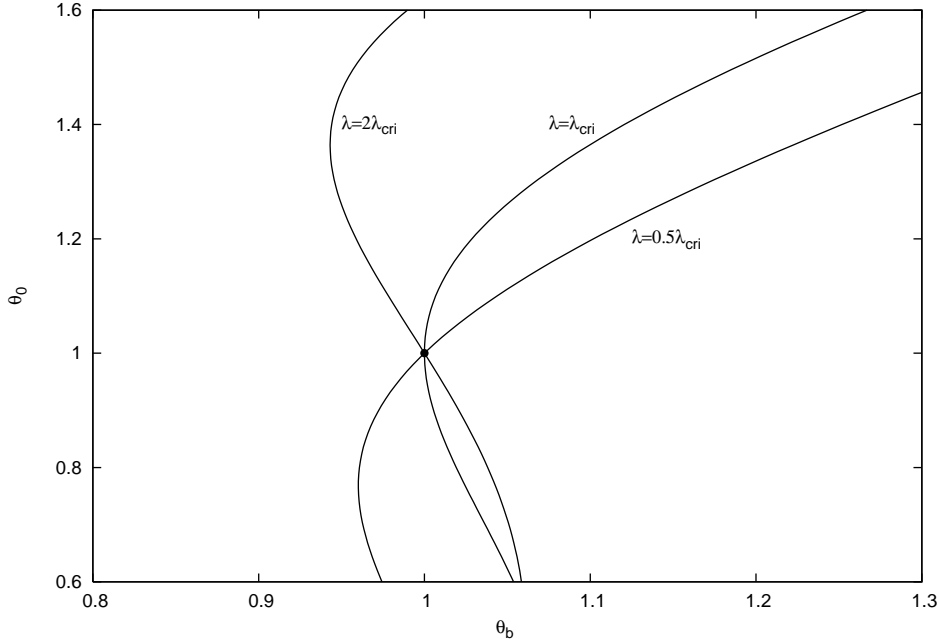}
\caption{\label{fig1} The central temperature ($\theta_0$) as
a function of the boundary temperature ($\theta_b$) with
$k=1/2$, $m=-1$, $n=-3/2$, for the case $\eta=0$ and the labeled
values of $\lambda$.}
\end{figure*}

When $\eta=0$ (no internal heating source) Eq.~\ref{estacionaria}
reduces to the case studied in Ref. \cite{Iba92}. In spite of
the simplicity of the case $\eta=0$, its study has already led
to a better understanding of the general problem of stability of
structures, even allowing to extend the study analytically up to
the second order \cite{Iba93}.
We see that for the case $\eta=0$ the trivial solution $\theta=1$,
i.e. the thermal equilibrium solution $T=T_{eq}$, is solution of
Eq.~\ref{estacionaria}. It has been shown that this solution is
linearly unstable if $\lambda$ fulfills the condition \cite{Iba92}:
\begin{equation}
\label{ec_criterio}
\lambda > \lambda_{cri} = \frac{\pi^2}{m-n} \ .
\end{equation}
The dimensionless parameter $\lambda$ (Eq.~\ref{lambda}) measures
the ratio between heating (or cooling) at equilibrium and thermal
difussion through the scale length $R_b$. Moreover, thermal
diffusion is a stabilizing factor because it tends to diminish
any temperature fluctuation. Thus, physically, the condition
$\lambda > \lambda_{cri}$ means that the generation of heat is
so large that it cannot be removed efficiently by thermal difussion,
and therefore a thermal instability develops. The analysis of the
rest of the stationary solutions (different from the trivial one)
cannot be done analytically but the stability can be inferred from
the position on the curve $\theta_0 \mathrm{\ vs.\ } \theta_b$.
In Fig.~\ref{fig1}
we show the plot resulting from integrating Eq.~\ref{estacionaria}
with $\eta=0$. The trivial solution $\theta_0 = \theta_b = 1$
(shown as a filled circle) is located on the positive slope branch
of the $\theta_0 \mathrm{\ vs.\ } \theta_b$ curve for the case
$\lambda < \lambda_{cri}$. This case corresponds to the stable
solution according to Eq.~\ref{ec_criterio}. On the contrary, the
trivial solution is on the negative slope branch for the case
$\lambda > \lambda_{cri}$ (unstable solution). Moreover, for
$\lambda = \lambda_{cri}$  (marginally stable solution) the
trivial solution separates the positive and negative slope
branches. According to Ref. \cite{Iba92} (see also Refs.
\cite{Iba91,Lan87,Fra55}), the positive slope branch can be
identified as the stable branch. Therefore, {\it stationary
solutions located on the positive (negative) slopes branches
are expected to be linearly stable (unstable)}. This is the
stability criterion we are using in this work.

For the more general case ($\eta \neq 0$) we see that $\theta=1$
is not a stationary solution of Eq.~\ref{estacionaria}. Both the
spatial dependence of the internal heating and its non-local
character (i.e., the local heating depends on material state in other
points through the optical depth) complicate the mathematical
treatment in such a way that it prevents us from finding an analytical
stationary solution. Thus, a stability analysis is not possible
but nevertheless we expect the above stability criterion to remain
valid. The argument for this claim comes from the relative behaviors
of thermal conduction and net heating/cooling with respect to the
temperature. When two stationary solutions are obtained for a
given boundary temperature (for example when $\theta_b > 1$ for
the case $\lambda=\lambda_{cri}$ in Figure~\ref{fig1}), only
one of them is stable (see Refs. \cite{Lan87,Fra55}). Let us
assume, for example, that both thermal conduction $Q(T)$ and
net heating rate $L(T)$ are increasing functions of temperature.
Two stationary solutions means that $Q=L$ at two temperature
values ($T=T_{eq}$). Necessarily, we have $dQ/dT > dL/dT$ at
one point and the opposite in the other. The point at which
$dQ/dT > dL/dT$ is stable because any small temperature
variation will be eventually damped by thermal conduction.
On the other hand, if the heating rate increases with
temperature faster than the thermal conduction then any
small temperature rise will grow. This argument should
remain true as long as the temperature dependences remain
the same. For a fixed position $r$, the dependence of the
internal heating on $T$ is the same as in the previous
papers ($\sim T^{-1}$, i.e. proportional to density).
This is the reason why we expect the stability criterion
to be valid for the case under study.

In any case, we performed some numerical tests in order
to check the validity of this criterion for the more
general and complicated case of a non-local heating
mechanism. For a given set of free parameters ($\lambda$
and $\eta$) we first find numerically the corresponding
stationary solutions for central temperatures around the
``turning point temperature", i.e. around the central
temperature value that separates the positive and negative
slope branches. Then we impose different temperature
perturbations (we use gaussian, exponential, and
sinusoidal profiles) around randomly chosen positions
in the structure. What we did was to check that
$\partial T / \partial t$ changes its sign exactly
at the turning point, in such a way that on the positive
slope branch the sign is opposite to the perturbation
(stable branch) and viceversa. We performed many tests
with different combinations of parameters and the results
obtained were the same. Notwithstanding this is not a
demonstration, these tests allow us to be much more
confident about the validity of the stability criterion
for the case under consideration.

\section{\label{sec_resultados} Effect of the internal heating}

To numerically solve Eq.~\ref{estacionaria} with the boundary
conditions~(\ref{frontera1}-\ref{frontera2}) we use the
Runge-Kutta method of order 4 with adaptive step size control
\cite{Pre92}. The distinctive feature of Eq.~\ref{estacionaria}
is the fact that to evaluate the internal heating term we need to
know the solution we are looking for.
The density is linked to the temperature (because of the isobaric
assumption) and therefore the integral in Eq.~\ref{tau} ultimately
depends on the temperature distribution from the source (as we
have indicated explicitly in Eq.~\ref{estacionaria}). Thus, we
have to proceed by successively iterating Eq.~\ref{estacionaria}
until $\theta (r)$ converges to the stationary distribution. Then,
given a central temperature $\theta_0$, in the first step we assume
an initial distribution $\theta (z) = \theta_0$ to obtain an initial
estimate of the optical depth $\tau(z)$. Then we use this result to
numerically integrate Eq.~\ref{estacionaria} and obtain a new
temperature profile. Now this temperature distribution is used
to recalculate $\tau(z)$, and so on. The initial conditions
are always satisfied, thus the central temperature remains
constant during these steps. The procedure is then repeated
until a convergence criterion is satisfied. In particular, we
require that temperature values computed at two successive
steps differ less than a prefixed small tolerance. Furthermore,
we require this condition to be met at several points inside the
structure. The final result is the whole temperature distribution
from the center to the boundary. The central temperature can then
be changed to search for other solutions with their corresponding
boundary temperatures.

The coefficients $\lambda$ and $\eta$ are determined by the physical
processes involved (pressure, external heating, cooling, thermal
conduction, chemical composition, etc.). Here we consider typical
ISM conditions. The mean pressure of the warm gas in the ISM is
$p/k_B \simeq 10^3\ K\ cm^{-3}$ \cite{Jen01}, which is heated
mainly (in absence of nearby stars) by the galactic radiation field.
Under these conditions, we have $\Gamma_0 = 2.7 \times 10^{-23}$,
$\Lambda_0 = 10^{-21}$, $m=-1$, and $n=-1.5$ \cite{Spi78,Mck77}
(in this work $\Gamma$ and $\Lambda$ are in units of
$erg\ cm^{-3}\ s^{-1}$).
This yields an equilibrium temperature $T_{eq} \simeq 1370\ K$.
Under these conditions, the thermal conduction is mainly due to
neutral particles and therefore $\kappa_0 = 2.5 \times 10^3$ and
$k=1/2$ \cite{Par53}. The cross section is assumed to be
$\sigma = 6.3 \times 10^{-18}\ cm^{2}$ \cite{Spi78}. We keep
as constants these fiducial values whereas the region size ($R_b$)
and the internal heating efficiency ($\gamma$) are treated as free
parameters. From Eqs.~\ref{lambda} and \ref{ec_criterio} we get
that the case in which the trivial solution is marginally stable
corresponds to structures with radius $R_{cri} \equiv \left(
\lambda_{cri}\,\kappa_0\ T_{eq}^{k-m+1} / \Gamma_0 \right)^{1/2}
\simeq 0.12\ pc$. The total optical depth (in thermal equilibrium)
would be $\tau_0 \simeq 1.7$. As mentioned before, $\lambda$
measures the ratio between net external heating at equilibrium
and thermal conduction through the structure. The trivial
solutions is stable when $\lambda < \lambda_{cri}$ because
thermal conduction smooths the energy gradient more efficiently
in relatively small structures ($R_b < R_{cri}$). On the other
hand, $\eta$ gives in some way the relative importance of the
internal heating and thermal conduction terms. Thus, given three
different mechanisms of energy transport (thermal conduction, net
external heating, and internal heating) there are only two free
parameters ($R_b$ and $\gamma$, related to $\lambda$ and $\eta$
through Eqs.~\ref{lambda} and \ref{eta}) determining the relative
importance among them.

\begin{figure*}
\includegraphics{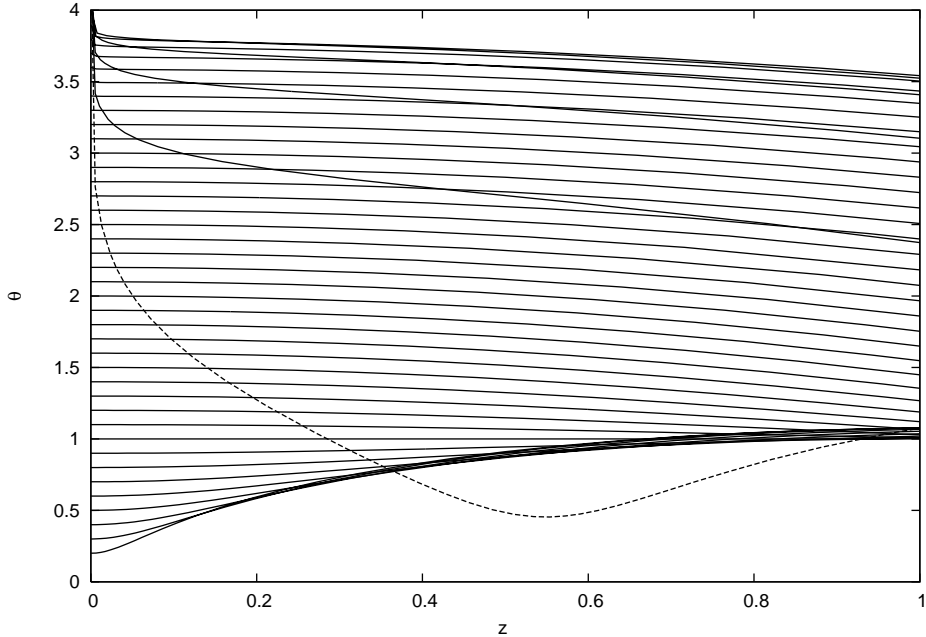}
\caption{\label{fig2} The distributions of temperature with
$k=1/2$, $m=-1$, $n=-3/2$, for the case $R_b=R_{cri}$
and $\gamma=10^{-3}$. Dashed line corresponds to the
highest central temperature shown ($\theta_0=4.4$).}
\end{figure*}

The temperature distributions inside the structure for the case
$R_b = R_{cri}$ and $\gamma = 10^{-3}$ \cite{Hol90} are shown
in Fig.~\ref{fig2}.
Each line in this figure is one solution $\theta(z)$ of
Eq.~\ref{estacionaria} for a given value of central
temperature $\theta_0$ and, therefore, of internal heating.
We see that there are no solutions with boundary
temperature ($\theta_b$) below $1$. Then, all the solutions
with $\theta_0 < 1$ have positive temperature gradients. As
we will see later, these solutions are thermally unstable.
For $\theta_0 > 1$, we have that $\theta_0$ increses as
$\theta_b$ increases, but for $\theta_0 \gtrsim 4$ this
behavior changes notoriously. For the highest central
temperature shown, we can see a sign change in the
temperature gradient around $z \simeq 0.55$. To
understand this behavior, we have calculated the relative
contribution of each term in Eq.~\ref{estacionaria} and
we have seen that at central temperatures $\theta_0
\lesssim 2$ the internal heating is always much smaller
than the (external)heating/cooling term. Thus, the
situation is almost the same as if $\gamma=0$, i.e.,
(external)heating/cooling balanced only by
thermal conduction. In contrast, at higher
$\theta_0$ values the internal heating dominates
over the (external)heating/cooling, but this
internal heating decreases very quickly with the
distance from the center and then a change in regime
occurs at some point (depending on the central temperature).
For the highest central temperature shown in Fig.~\ref{fig2},
this change in regime occurs precisely at $z \simeq 0.55$.
We will see in next figure that this central temperature is
very close to the maximum central temperature for which
there exists a solution. Above this maximum, the temperature
gradient becomes so large that the temperature goes down
to unphysical negative values (i.e. there is no physical
solution). For central temperatures between $\sim 2$ and
this maximum temperature the internal heating term always dominates at
$z \lesssim 0.5$.
Therefore, the influence of the radiation source seems to be
important only around the central region (for the values of
constants and parameters considered here), but its effect
can be enough to make the system unstable, as we will show
next.

\begin{figure*}
\includegraphics{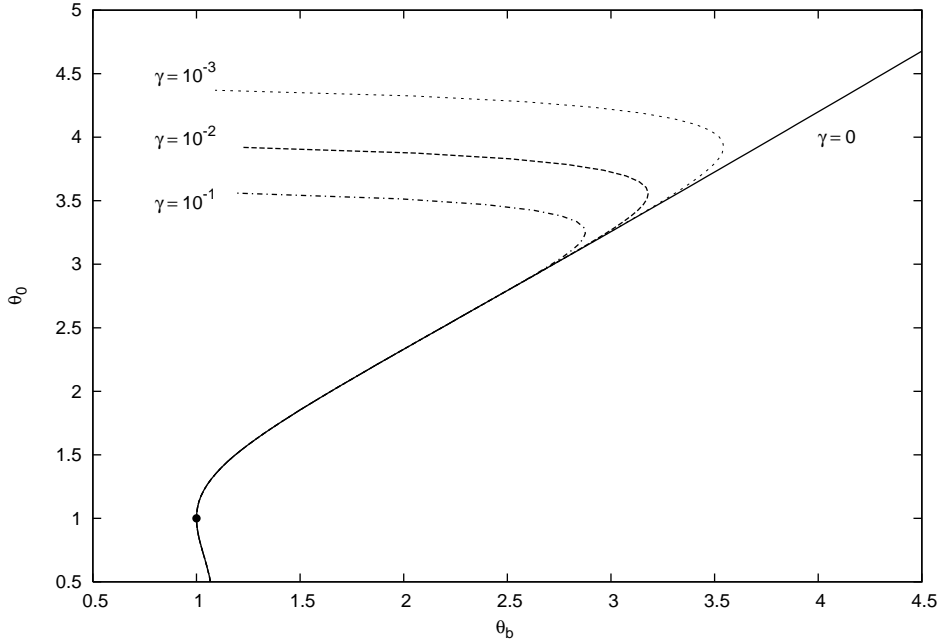}
\caption{\label{fig3} The central temperature ($\theta_0$) as
a function of the boundary temperature ($\theta_b$) with $k=
-1/2$, $m=-1$, $n=-3/2$, for the case $R_b=R_{cri}$ and
the labeled values of $\gamma$. The thermal equilibrium
solution is indicated by a filled circle.}
\end{figure*}

The $\theta_0 \mathrm{\ vs.\ } \theta_b$ diagram for $R_b = R_{cri}$
and various values of $\gamma$ is shown in Fig.~\ref{fig3}.
Each point in this figure represents one solution of
Eq.~\ref{estacionaria} (i.e., one line in
Fig.~\ref{fig2}), and each line represents a family of
possible solutions for each pair of values ($R_b$, $\gamma$).
Note that all curves cross through the trivial solution
($\theta_0 = \theta_b = 1$, shown as a filled circle).
This is contrary to the expected behavior because $\theta =
1 = constant$ is not a solution of the energy equation. But,
as mentioned before, at central temperatures around $\theta_0=1$
the internal heating can be neglected and the situation is
equivalent to the case $\gamma=0$. According to the stability
criterion (Sec.~\ref{sec_criterio}), the solutions located on
the positive slope branch are stable, and viceversa. For the
case $\gamma = 0$ we recover the solutions studied in Ref.
\cite{Iba92}, where the trivial solution is marginally
stable and the other solutions with $\theta_0 > 1$ are stable.
For the cases $\gamma \neq 0$ we observe that there exists
a maximum central temperature (hereinafter $\theta_{0,max}$) above
which the solutions become thermally unstable. For the case
$\gamma=10^{-3}$ this threshold occurs at $\theta_{0,max} \simeq
3.9$, but clearly $\theta_{0,max}$ decreases as $\gamma$ increases.

\begin{figure*}
\includegraphics{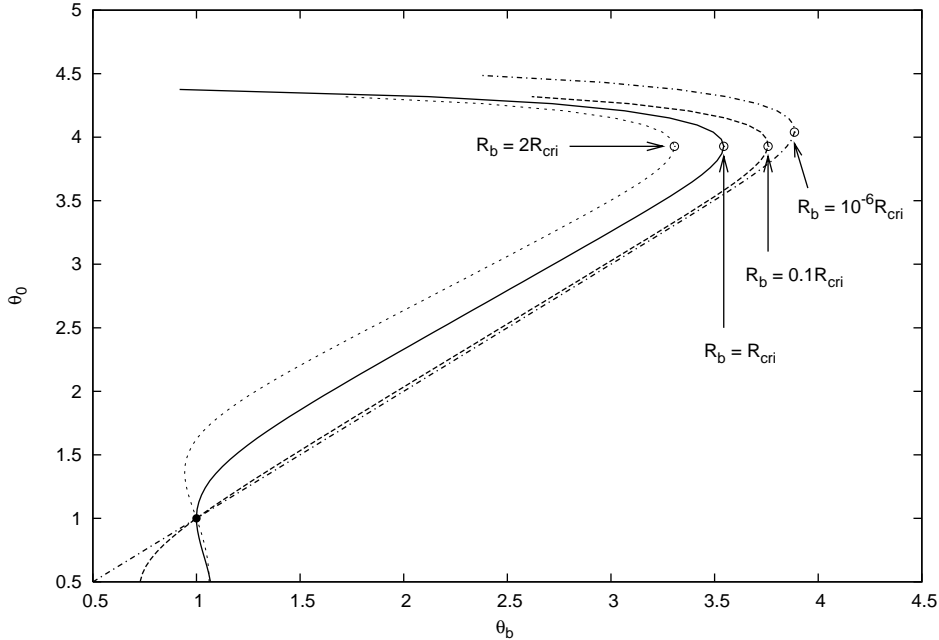}
\caption{\label{fig4} As in Fig.~\ref{fig3} but for the case
$\gamma = 10^{-3}$ and the labeled values of $R_b$. The thermal
equilibrium solution is marked by a filled circle, and the open
circles indicate the threshold solution $\theta = \theta_{0,max}$.}
\end{figure*}

In Fig.~\ref{fig4}
we have fixed the internal heating efficiency ($\gamma = 10^{-3}$)
and we have varied the radius $R_b$. At low central temperatures
the behavior is similar to the case $\gamma = 0$, i.e., the trivial
solution is stable for  $R_b < R_{cri}$ and unstable for
$R_b > R_{cri}$. At relatively high central temperatures, we
obtain that the threshold temperature $\theta_{0,max}$ for thermal
instability (indicated as open circles in Fig.~\ref{fig4}) is
approximately constant. This interesting result is a direct
consequence of the ``local" character of the internal heating:
this heating mechanism decreases too quickly with $z$ that its
destabilizing effect does not depend on the structure size. It
can be seen that there is a small variation in $\theta_{0,max}$
for the extreme case $R_b = 10^{-6} R_{cri}$, but this value is
so small ($\sim 5 R_{\rm sun}$ for our fiducial values) that
its validity is questionable.

In the case of a star that migrates into a diffuse interstellar
region, the quantity $\theta_{0,max}$ would represent the photospheric
(effective) temperature of a star above which the surrounding
medium becomes thermally unstable. We have to point out that
this simple interpretation is possible only if the time-scale
for thermal conduction is shorter than any other time scale
(e.g. the dynamical time) that could be involved in the problem.
For our fiducial case with $\gamma = 10^{-3}$ and $R_b = R_{cri}
\simeq 0.12\ pc$ we obtain $\theta_{0,max} \simeq 3.9$.
An interesting result is that $\theta_{0,max}$ does not depend on,
or depends very weakly on, $R_b$ (at least under the conditions
and assumptions considered in this work). In other words,
our results suggest that a solar-like star with surface
temperature $\sim 5300\ K$ may possibly destabilize the medium
in its vecinity, almost independently of the region size.
After this, the surrounding medium will undergo
a transition to another state (or phase). Differences
in the chemical composition and/or the amount of dust from
region to region in the Galaxy (or in other galaxies) may
change the value of $\gamma$ \cite{Hol90}, and consequently
the maximum central temperature. Here we obtain $\theta_{0,max}
= 3.6$ for $\gamma = 10^{-2}$ and $\theta_{0,max} = 3.2$ for
$\gamma = 10^{-1}$.

If stars eventually migrate into interstellar regions, then
thermal instabilities caused by these stars may play a
non-negligible role in stimulating phase transitions
in the ISM. However, the present model
is clearly too simple to be directly applicable to the ISM.
The strongest restriction is perhaps
isobaricity, which is a difficult condition to be fulfilled
even in the diffuse ISM \cite{Jen01}. Notwithstanding these
limitations, our results suggest that thermal instabilities
driven by stars could be a key mechanism in the ISM.


\section{\label{sec_conclusiones} Conclusions}

We have studied the equilibrium and stability of isobaric,
spherical structures with an internal radiation source
located at its center. We find that there exists a threshold
central temperature above which the structure becomes
thermally unstable, and this temperature decreases as
the internal heating efficiency increases. Interestingly,
the threshold temperature does not depend on the region
size, because the destabilizing mechanism has only a
local effect. We have shown that thermal instability
triggered by an embedded source may have interesting
consequences in the ISM evolution which should be
addressed in future, more complete, studies.

\begin{acknowledgments}
We are very grateful to the anonymous referee for the
critical and constructive report, which improved this
paper. N.~S. acknowledges financial support from MEC
of Spain through grant AYA2007-64052.
\end{acknowledgments}

\end{document}